# Topological defects and geometric memory across the nematic-smectic A liquid crystal phase transition

Ahram Suh[1], Min-Jun Gim[1], Daniel Beller[2,*] and Dong Ki Yoon[1,3,†]

[1]Graduate School of Nanoscience and Technology, KAIST, Daejeon, 34141, Republic of Korea

[2]Department of Physics, University of California, Merced, CA 95343, USA

[3]Department of Chemistry and KINC, KAIST, Daejeon, 34141, Republic of Korea

**Abstract**

**We study transformations of self-organized defect arrays at the nematic-smectic A liquid crystal phase transition, and show that these defect configurations are correlated, or "remembered", across the phase transition. A thin film of thermotropic liquid crystal is subjected to hybrid anchoring by an air interface and a water substrate, and viewed under polarized optical microscopy. Upon heating from smectic-A to nematic, a packing of focal conic domains melts into a dense array of boojums—nematic surface defects—which then coarsens by pair-annihilation. With the aid of Landau-de Gennes numerical modeling, we elucidate the topological and geometrical rules underlying this transformation. In the transition from nematic to smectic-A, we show that focal conic domain packings are organized over large scales in patterns that retain a geometric memory of the nematic boojum configuration, which can be recovered with remarkable fidelity.**

[*] dbeller@ucmerced.edu
[†] nandk@kaist.ac.kr

## Introduction

Topological defects are localized, quantized breakdowns of material order, and their character is determined by the symmetries of the ordered medium [1-3]. Understanding their behavior is essential for modeling material properties such as plastic deformation and phase transitions, and a great deal of research in soft matter focuses on harnessing topological defects for self-assembly [4-10]. Liquid crystals (LCs) provide an important arena for the controlled creation of topological defects by geometric frustration, and for the optical observation of these defects at the micron scale [11-16]. Many LC materials present multiple liquid crystalline phases, with distinct symmetries, as temperature or concentration is varied [17].

In the phase transition between any two ordered phases with different symmetries, the character of topological defects generally changes. However, much remains to be understood about whether and how the *configurations* of defects may be correlated across the phase transition. Recently, a simple experiment studied the nematic (N)-to-smectic A (SmA) LC phase transition in the material, 8CB (4'-octyl-4-biphenylcarbonitrile) on a water substrate [18]. It was demonstrated that boojums, surface singularities in the N phase, may determine the subsequent packing geometries of focal conic domains (FCDs), defect motifs of the SmA phase.

In this work, we demonstrate that spontaneously assembled *patterns* of boojums in the N phase determine the large-area defect patterns that arise in the SmA phase, and vice versa, through newly observed kinetic pathways in the form of geometrical and topological transformations. While N boojums and Sm FCDs create distortions that propagate into the bulk, they are topological defects only at the boundary. As shown in the previous work [18], we study floating films of the LC material, 8CB on a water substrate, with a free air interface above. Strong homeotropic (normal) anchoring at the air interface and degenerate planar (tangential) anchoring at the water interface combine to create hybrid anchoring conditions, which introduce a geometrical frustration known to be relaxed by boojums [19] and FCDs [20].

Even though the self-assembled defect patterns in the two phases are quite different in symmetry and length scale, we identify morphological pathways connecting the two configurations across the phase transition. Within a narrow temperature interval below $T_{NA}$, the Sm FCDs carry a geometric memory of the N boojum configuration, which can be recovered with remarkable fidelity upon heating. Once this geometric memory is lost, we show that a packing of FCDs can be melted into a dense network of boojums, which we relate to the FCDs' planar graph with the support of Landau-de Gennes numerical modeling. A subsequent rapid coarsening of the boojums by pair annihilation establishes a new stable N boojum configuration.

**Defect configurations**

Water provides an ideal substrate material to avoid any surface-imposed preference of in-plane direction and to ensure uniform LC film thickness. In addition, LC films of micron-scale thickness on such substrates are ideal for studying generation of LC defects, which arise due to the competition between elastic energy (Frank constant $K \sim 10^{-11}$ N) and surface anchoring energy (anchoring strength $W \sim 10^{-5}$ J/m$^2$): the ratio of these two constants gives a characteristic length scale called the de Gennes-Kleman extrapolation length, $L_d = K/W \sim 10^{-6}$ m. The droplet of 8CB is deposited onto a water substrate in the temperature range of 8CB's isotropic phase. The droplet spreads until reaching a uniform thickness depending on the volume of the LC sample.

Upon cooling of the LC material into the N phase, hybrid alignment causes a distorted N director $\boldsymbol{n}$ field to emerge, rotating from vertical at the air interface to strongly tilted into the horizontal at the water interface. Because the anchoring at the water interface is *degenerate* planar, the choice of tilt direction $\boldsymbol{c}$ in the *xy* plane (i.e., normalized horizontal projection) is a spontaneous symmetry breaking. Domains of common tilt direction grow, but at certain points on the substrate, $\boldsymbol{c}$ winds through an integer multiple of $2\pi$. These are the boojums, topological surface defects with integer winding number *s*. Away from the surface, singularities in the director field are avoided through escape in the third dimension [21]. Under polarized optical microscopy (POM), boojums are marked by the intersection of $4|s|$ dark brushes (Figure 1a,d,g). As all boojums observed in this study have four dark brushes, they all have $s=\pm 1$. Positive (Fig. 1a,d) and negative (Fig. 1g) values are distinguished by rotation of the dark brushes respectively with, or counter to, rotation of the crossed polarizers.

The boojums self-organize into a square lattice, albeit one with many dislocations and large variations in spacing (Fig. 2a; see also Fig. 3f,g and Supplemental Material, Fig. S1a,b [22]). While the presence of boojums might appear to indicate incomplete relaxation of the N directors, in fact these defects are equilibrium features. De Gennes [23] predicted that antagonistic forces on the NLC, such as those supplied by hybrid anchoring conditions, could cause square arrays of surface defects associated with slight spatial variations of the height of the planar anchoring interface. These defects were observed by Meyer [24] and shown to form square lattices by Madhusudana and Sumathy [19]. Their equilibrium spacing scales with the capillary length, which for the 8CB-water interface is ~2.1 mm, greater than $L_d$ [25].

The N defect lattices contain three distinct types of $|s|=1$ boojums. Two of these, the +1 converging and +1 diverging boojums, have a radial $\boldsymbol{c}$ configuration at the water substrate (Fig. 1a,d and Fig. 1c,f). Assigning an upward orientation to *n* at the air interface (this choice is arbitrary but can be made consistently [18]), *n* and *c* are oriented radially inward around the +1 converging ($+1_C$) boojum and radially outward around the +1 diverging ($+1_D$) boojum. The third type is the −1 boojum (Fig. 1g, Fig. 1i), which in our convention can be considered converging along one axis and diverging along the perpendicular axis. The −1 boojum's hyperbolic cross-section is responsible for the square symmetry of the lattice. The necessity of all three types of boojums is seen immediately, for example, upon mapping the boojum configuration in Fig. 2a to a "nail-head" diagram indicating the tilt of *n* in Fig. 2d. Not only do +1 and −1 boojums alternate on neighboring lattice sites, but in addition, neighboring rows differ in having $+1_C$ or $+1_D$ boojums. Having only one type of +1 boojum would necessitate grain boundary discontinuities or zeros of the *c* vector field.

In the SmA phase, consisting of single molecule-thick layers, the LC molecules subjected to hybrid anchoring form a self-assembled packing of FCDs. In each FCD, the layers are curved with their centers of curvature lying along an ellipse of eccentricity *e*, at the LC-water interface, and a branch of a hyperbola, rising into the bulk from one of the ellipse's foci (Fig. 2e) [26]. Under POM, each FCD is visible as a set of bright lobes separated by dark bands, all meeting at the ellipse focus (Fig. 1b,e,h). A special case with rotational symmetry, i.e. four equal-sized bright lobes, is seen in the center of Fig. 1b. This is the toric FCD (TFCD), in which the ellipse is a circle (eccentricity: *e*=0) and the hyperbola is a vertical straight line through the circle's center. While the hyperbola is a cusp defect rather than a topological defect, the layer normal direction (i.e., the director field) has a +2π winding about the hyperbola at the ellipse's focus on the LC-water interface (see Fig. 2f).

As recently demonstrated in Ref. [18], the character of an individual boojum determines the orientations of FCDs that initially form nearby, when the NLC is cooled into the SmA phase. An FCD's orientation $u$ is taken to be the direction into which the hyperbola bends with increasing height, parallel to the long axis of the ellipse. The FCDs are oriented radially inward toward the site of a $+1_C$ boojum (Fig. 1b), radially outward from the site of a $+1_D$ boojum (Fig. 1e), and in a diamond pattern around the site of a $-1$ boojum (Fig. 1h). As explained in Ref. [18], the formation of FCDs in an already distorted director field favors $u \approx c$. Furthermore, the FCDs self-organize in rows parallel to $u$. In the $+1_C$ case, but not the others, at the site of the boojum there arises a TFCD, for which the in-plane orientation $u$ is undefined, although $u$ is oriented radially inward nearby.

Consequently, the FCD configuration that arises upon cooling through the phase transition bears the geometrical imprint of the boojum configuration over the scale of hundreds of microns. As shown in Fig. 2b, the Sm develops a complex checkerboard of FCDs rows oriented along the diagonals of the boojum lattice. The points of convergence, previously the sites of the $+1_C$ boojums, each have a TFCD. Grain boundaries in the orientations of FCD rows lie along the lines connecting neighboring boojums. We can draw a square unit cell in which all FCDs are oriented toward the center; then the square's vertices are the $+1_D$ boojum sites, and the $-1$ boojum sites are near the midpoint of each edge. With further cooling (~32℃), the TFCDs become more distinct (Fig. 2c).

Upon still further cooling (~31℃), the FCDs with $e \neq 0$ lose their eccentricity and become TFCDs (Fig. 3a) [18]. This loss of eccentricity has an important consequence for the Sm phase's geometric memory of the N boojum configuration, because the field of tilt directions $c$ is completely erased. If the LC is cooled into the SmA phase but kept above the temperature where all FCDs become TFCDs, then the sample can be heated back into the N phase with the boojums in almost exactly the original configuration (Fig. 4, Supplemental Material, Fig. S1 [22]). This reversibility is striking given that the boojum configuration is nearly unrecognizable by eye in the FCD packing. On the other hand, cooling the FCDs until they become TFCDs erases the geometric memory: subsequent heating into the N phase establishes a completely new boojum configuration (see Supplemental Material, Fig. S1 [22]), by a process that we describe next.

The pathway by which a new boojum configuration is established from the melting of a TFCD packing is quite distinct from the transformations observed during cooling. Whereas cooling gives rise to many FCDs per boojum, heating transforms *each* TFCD into a boojum; thus a dense array of boojums arises from the TFCD packing. Figure 3b shows this transformation, with the upper right portion of the image in the N phase and the bottom left portion still in the SmA phase. At the line where the transition is occurring, POM no longer shows a clear distinction between boojums and TFCDs: each TFCD in the SmA phase becomes a boojum in the N phase. As highlighted in the yellow boxed region in Fig. 3b, the quasi-hexagonal packing of the Sm TFCDs gives way, at the transition line, to the square arrangement favored by a N boojum network. Subsequently, the boojums coarsen rapidly by pair annihilation, as seen in the sequence of images Fig. 3b-d, until the spacing between defects has grown from a few microns in the SmA phase to a few tens of microns in the N phase.

The pair annihilation of nearly all the boojums requires, of course, $s=+1$ and $s=-1$ boojums in almost equal numbers. However, the TFCDs supply only the $+1$ boojums: As illustrated in Fig. 2f, the director field at the base of each TFCD matches that of a radial $+1$ boojum. More specifically, the vertical cross-section of the director field in Fig. 2f shows that the TFCD is a $+1_C$ boojum. As discussed above, in order to produce a smooth director field, each $+1_C$ boojum needs not only a partner $-1$ boojum, but also a $+1_D$ boojum with its own partner $-1$ boojum—a total of three other boojums for each TFCD. Where do these other boojums come from?

The answer is found in the return of a horizontal component *c* to the director field in the interstices between FCDs, as the SmA phase is heated into the N phase. Deep in the SmA phase, the interstices between TFCDs are filled by planar horizontal layers, with vertical *n*. This configuration locally incurs the maximum possible penalty of energy per unit area, $W_1$, from the water interface's degenerate planar anchoring potential, a situation that can only occur if the elastic energy cost of distortions outweighs the surface anchoring potential. As the N phase is cooled into the SmA phase, the bend elastic constant $K_3$ diverges, and the director field in the interstices becomes completely vertical when $K_3$ exceeds the product *HW* of the film thickness *H* and the anchoring strength *W* [27,28]. Conversely, as the SmA phase is heated into the N phase, $K_3$ decreases and eventually *W* is strong enough to induce a tilt of the director field off of the vertical.

The substrate anchoring potential, being degenerate planar, imposes no preference on the direction *c* of this tilt into the horizontal plane. Therefore *c* would be a spontaneously broken symmetry, were it not for the fact that the director field is already tilted nearby within the TFCDs. We therefore expect the tilt direction to vary spatially, and to be decided locally by the tilt in the nearest portion of a TFCD, so as to decrease the elastic energy cost of the tilt. In the idealized TFCD packing illustrated in Fig. 5a, nail-head markings show the extension of *c* from within the TFCDs to the interstitial regions nearby. Simply by requiring *c* to be nonzero and continuous almost everywhere, we are led to two predictions: A $-1$ boojum forms at the intersection of every pair of adjacent TFCDs, and a single $+1_D$ boojum forms in each interstitial region.

We can relate these boojums to the planar graph associated with the TFCD packing, with a vertex at each TFCD center and an edge connecting each pair of adjacent TFCDs. When tilt is restored to the interstices, there will be one −1 boojum along each edge of the graph, and one $+1_D$ boojum in each face. The total boojum charge is then $V–E+F–1$, where $V$ is the number of vertices, $E$ is the number of edges, and $F$ is the number of faces in the graph including the external face. This is one less than the formula for the Euler characteristic χ, whose value is 2 for any planar graph. Therefore, any isolated cluster of connected TFCDs has a total boojum charge of +1, regardless of the configuration of TFCDs. It follows that bringing into contact two connected clusters of TFCDs introduces a new −1 boojum at the new point of tangency so that the total charge of the new cluster is +1.

We observe such a transformation using POM with a first-order retardation plate (λ= 530 nm), as shown in Fig. 3e-g. In these images, cyan blue regions have director tilt *c* along the direction labeled λ, yellow regions have *c* in the perpendicular direction, and magenta regions occur wherever the director is vertical or is aligned with the polarizer or analyzer. In the SmA phase, the TFCDs (Fig. 3e) appear as two pairs of alternating cyan blue and yellow lobes, separated by dark lines meeting at four-fold crosses at the TFCD centers, marked by purple dots. Interstices between the TFCDs appear magenta because the director field is vertical. Just after melting to the N phase (Fig. 3f,g), the cyan blue and yellow lobes expand outward from each TFCD, indicating that the interstices adopt the tilt direction of the nearby region of TFCD. New four-fold dark crosses, indicating boojums of winding number ±1, are created where the cyan blue and yellow lobes from different TFCDs come into contact. For each pair of adjacent TFCDs, a −1 boojum forms at the point of tangency, marked by a light blue dot in Fig. 3f. A pink dot marks the $+1_D$ boojum that forms in each interstitial region. It appears that the $+1_C$ boojum from the TFCD at bottom left in Fig 3e has already annihilated with a $-1$ boojum at the time of Fig. 3f.

We find support for our hypothesis, regarding the genesis of the various boojum types, in the results of Landau-de Gennes (LdG) numerical modeling. We use this technique to study the free energy relaxation pathway of a N configuration from an initial condition whose director field matches that of a packing of TFCDs, with purely vertical orientation outside of the TFCDs. For example, Fig. 5b shows an initial condition whose director field is that of a square packing of four TFCDs. Homogeneous uniaxial order is imposed initially. Homoetropic anchoring is imposed at the top surface, and degenerate planar anchoring at the bottom surface, both with strong but finite anchoring strengths; free boundary conditions are employed in the horizontal directions. While the LdG model does not include Sm order, we study the N phase near $T_{NA}$ by setting the bend elastic constant significantly higher than the splay elastic constant, $K_3/K_1 = 8$.

The relaxation pathway during LdG numerical free-energy minimization provides a reasonable approximation of the relaxation of the physical system, despite neglecting hydrodynamics [29]. Boojums appear very early in the relaxation. As shown in Fig. 5c, a $+1_C$ boojum forms at the center of each TFCD, a $+1_D$ boojum forms at the center of the square, and a $-1$ boojum appears at each tangency point of two TFCDs, as predicted. The $+1_C$ and $-1$ boojums take the form of short $\pm\frac{1}{2}$ disclination line defects, whose two intersections with the bottom surface each have half of the total winding number. (This "split-core" structure has been observed in previous LdG numerical studies of boojums [30].) With further relaxation, the boojums move inward under the attractive influence of their neighbors with opposite winding number, eventually leading to pair annihilation of all boojums but one, of type $+1_C$. When the initial condition is a close-packed triangular configuration of three FCDs rather than a square of four, two of the three expected $-1$ boojums are combined with the central $+1_D$ boojum to make a large $-½$ disclination line. The smaller size scale of the numerics compared to experiment likely favors the disclination line configuration; the simulated TFCD radius is 288 nm, compared to ~5 µm in experiment.

**Conclusion**

In this work, we have demonstrated complex and distinct pathways for transformations of topological defects upon heating or cooling through the N-SmA phase transition, in a LC film with geometrical frustration due to hybrid anchoring. Boojums in the N phase self-organize into square lattices that leave their geometric imprint in the much denser packing of focal conic domains in the SmA phase: FCDs are oriented (tilted) toward the centers of square unit cells ($+1_C$ boojum sites) and away from the vertices ($+1_D$ boojum sites), with $-1$ boojum sites along the edges. The FCD packing retains, through the field of tilt directions, a geometric memory of the boojum configuration, such that re-heating allows recovery of the original N configuration. This geometric memory is lost when further cooling transforms the FCDs into TFCDs. In that case, subsequent melting creates a dense lattice of boojums at the TFCD packing's vertices, edges, and faces, and pair-annihilation of most of these boojums leads to a new equilibrium N configuration. Numerical modeling of the N phase using a Sm-like initial condition supports our hypothesis for the genesis of the three boojum types.

Our findings suggest that the N-SmA transition in hybrid-aligned LCs could be useful for the reversible aggregation and dispersion of inclusions—nanoparticles or colloidal particles—which are attracted to LC defects [31]. As the phase transition alters the defect lattice type and changes the defect spacing from microns to tens of microns, such composite systems could find applications in patterning tools as well as photonics [32,33]. If +1 boojum sites can be controlled, the interchange of $+1_C$ and $+1_D$ boojums would potentially introduce a complex multistability to both phases. Sculpting the shape of the fluid interface through capillarity has been shown to produce spatially heterogeneous FCD configurations [34-36], and could have profound implications for the phase transition here. As in Ref. [18], there is still a need for a detailed geometrical model and energetic analysis of the transition from $e \neq 0$ FCD rows to TFCDs with decreasing temperature. The dynamics of boojum pair-annihilation is also of interest for future study, as subtleties may arise from the aforementioned incompatibility of neighboring $+1_C$ and $+1_D$ boojums, as well as the possibility of ±½ disclination lines. More broadly, our findings in this simple setup encourage investigation into correlated topological defect configurations, transformation pathways, and geometric memory and hysteresis wherever a material has a transition between distinct ordered phases.

**Acknowledgements**

This work was supported by a grant from the National Research Foundation (NRF), funded by the Korean Government (MSIT) (2017R1E1A1A01072798 and 2017M3C1A3013923). D.A.B. thanks R. Pelcovits and T. Powers for partial support through National Science Foundation Grant Nos. MRSEC-1420382 and CBET-1437195.

**Author contributions**

A.S. and M.-J.G. contributed equally to this work.

**APPENDIX: EXPERIMENTAL AND NUMERICAL DETAILS**

*Sample preparation:* We used 4'-octyl-4-cyanobiphenyl (8CB) (purchased from Sigma-Aldrich) which has the N and SmA phase transition temperatures of 39.1℃ and 32.2℃, respectively. We dropped 8CB onto a water substrate in the temperature range of 8CB's isotropic phase. The phase transition was observed by controlling the temperature on a heating stage with a temperature controller (LTS420 and TMS94, LINKAM)

*Characterization*: All POM images were obtained using POM with and without a retardation plate (LV100POL, Nikon).

*Numerical Landau-de Gennes relaxation method:* The N configuration is numerically modeled by the traceless, symmetric, rank-3 tensor $Q(\mathbf{r})$ which, in a uniaxial N, is related to the director $\mathbf{n}$ by $Q_{ij} = \frac{3}{2} S \left( n_i n_j - \frac{1}{3} \delta_{ij} \right)$, where $S$ is the N degree of order. With $Q(\mathbf{r})$ defined on a regular cubic mesh, the following nondimensionalized Landau-de Gennes (LdG) free energy is minimized over $Q(\mathbf{r})$:

$$F_{\text{LdG}} = \int d^3\mathbf{r} \left( f_{\text{phase}} + \tilde{L} f_{\text{elastic}} \right) + \int dS\, f_{\text{anch}}$$

Here, the first integral in $F_{\text{LdG}}$ integrates the phase and elastic free energy densities over the bulk, while the second integral contains the surface anchoring contribution from the bottom (water) and top (air) interfaces. The bulk free energy densities are:

$$f_{\text{phase}} = -\frac{1}{2} \text{tr}(Q^2) + \frac{1}{3} \tilde{B} \text{tr}(Q^3) + \frac{1}{4} \tilde{C} \left( \text{tr}(Q^2) \right)^2$$

$$f_{\text{elastic}} = \frac{1}{2} \frac{\partial Q_{ij}}{\partial x_k} \frac{\partial Q_{ij}}{\partial x_k} + \frac{1}{2} \ell_2 \frac{\partial Q_{ij}}{\partial x_j} \frac{\partial Q_{ik}}{\partial x_k} + \frac{1}{2} \ell_3 Q_{ij} \frac{\partial Q_{kl}}{\partial x_i} \frac{\partial Q_{kl}}{\partial x_j}$$

The "phase" free energy parameters $\tilde{B}$ and $\tilde{C}$ are respectively taken to be -12.3 and 10.1, values commonly assumed in modelling 5CB [29], giving preferred bulk degree of order $S = S_0 = 0.533$. $\tilde{L}$ is a dimensionless parameter controlling the importance of the elastic free energy compared to the phase free energy. We set this to $\tilde{L} = 2.32$; following [29], the mesh spacing is then 4.5 nm. The Frank elastic constant ratios $k_i = K_i/K_1$ are related to the Q-tensor elastic constant ratios $\ell_i$ by $\ell_2 = -6(k_2 - 1)/(k_3 + 3k_2 - 1)$, $\ell_3 = (2/S_0)(k_3 - 1)/(k_3 + 3k_2 - 1)$. We keep $k_2 = 1$, $k_3 = 8$.

In the anchoring energy density $f_{\text{anch}}$, degenerate planar anchoring at the water interface is modelled using the orientational anchoring component of Fournier and Galatola's anchoring potential [37], $f_{\text{anch}} = W_1(\tilde{Q}_{\alpha\beta} - \tilde{Q}^\perp_{\alpha\beta})(\tilde{Q}_{\alpha\beta} - \tilde{Q}^\perp_{\alpha\beta})$, where $\tilde{Q}_{\alpha\beta} = Q_{\alpha\beta} + \frac{1}{2}S_0\delta_{\alpha\beta}$, $\tilde{Q}^\perp_{\alpha\beta} = P_{\alpha\gamma}\tilde{Q}_{\gamma\delta}P_{\delta\beta}$, using the projection operator $P_{\alpha\beta} = \delta_{\alpha\beta} - \nu_\alpha\nu_\beta$ with substrate normal $\hat{\nu} = \hat{z}$. Homeotropic anchoring at the air interface is modeled using the Nobili-Durand anchoring potential $f_{\text{anch}} = W_0(Q_{\alpha\beta} - Q^0_{\alpha\beta})(Q_{\alpha\beta} - Q^0_{\alpha\beta})$, where $Q^0_{\alpha\beta} = \frac{3}{2}S_0\left(\nu_i\nu_j - \frac{1}{3}\delta_{ij}\right)$. We use the values $W_0 = W_1 = 5.82$, so that anchoring at both surfaces is in the strong anchoring regime. Torque-free boundary conditions are at the $x$ and $y$ boundaries to simulate an isolated TFCD cluster (as opposed to laterally confined or periodic).

$F_{\text{LdG}}$ is minimized using a finite difference scheme on a regular cubic mesh, with a conjugate gradient algorithm from the ALGLIB package (http://www.alglib.net). The initial conditions consist of the director field defined by the layer normals of TFCDs (oriented at any point $P$ along the line through $P$ connecting the TFCD's circular base to its central line defect, and oriented vertically in the interstices between TFCDs), with uniaxial order and $S = S_0$ everywhere, plus small random perturbations.

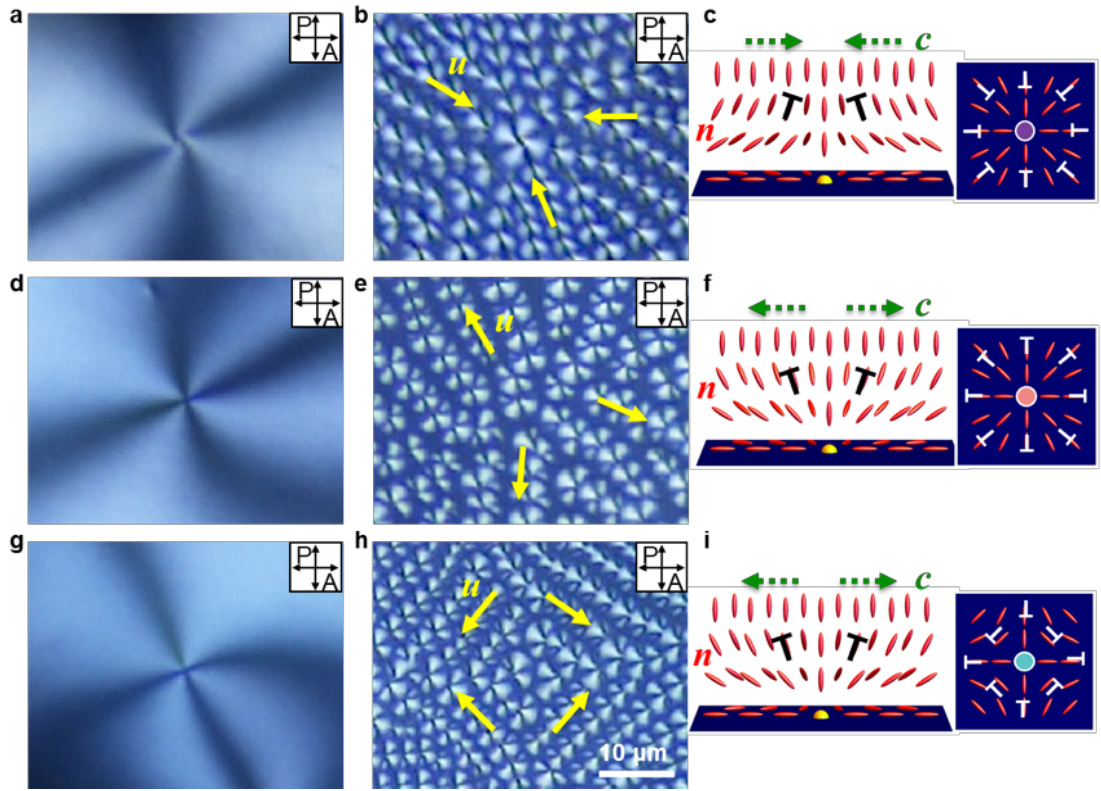

FIG. 1. Configuration of the LC orientation in each boojum defect based on N to SmA phase transition. (a,d,g) POM images of each NLC boojum defect; (a) +1 converging boojum ($+1_C$), (d) +1 diverging boojum ($+1_D$), (g) -1 boojum. (b,e,h) POM images of FCDs assembled at the sites of each boojum defect; (b) $+1_C$ boojum, (e) $+1_D$ boojum, (h) -1 boojum. Yellow arrows indicate local FCD orientation *u*. (c,f,i) Schematic diagrams of director field *n* at each type of boojum, viewed in cross-section from the side (left) and in the plane of the substrate from above (right); (c) purple dot ($+1_C$ boojum), (f) pink dot ($+1_D$ boojum), and (i) light blue dot (-1 boojum). T-shaped nail-heads indicate tilt of *n*; the horizontal projection *c* is indicated by green dashed arrows.

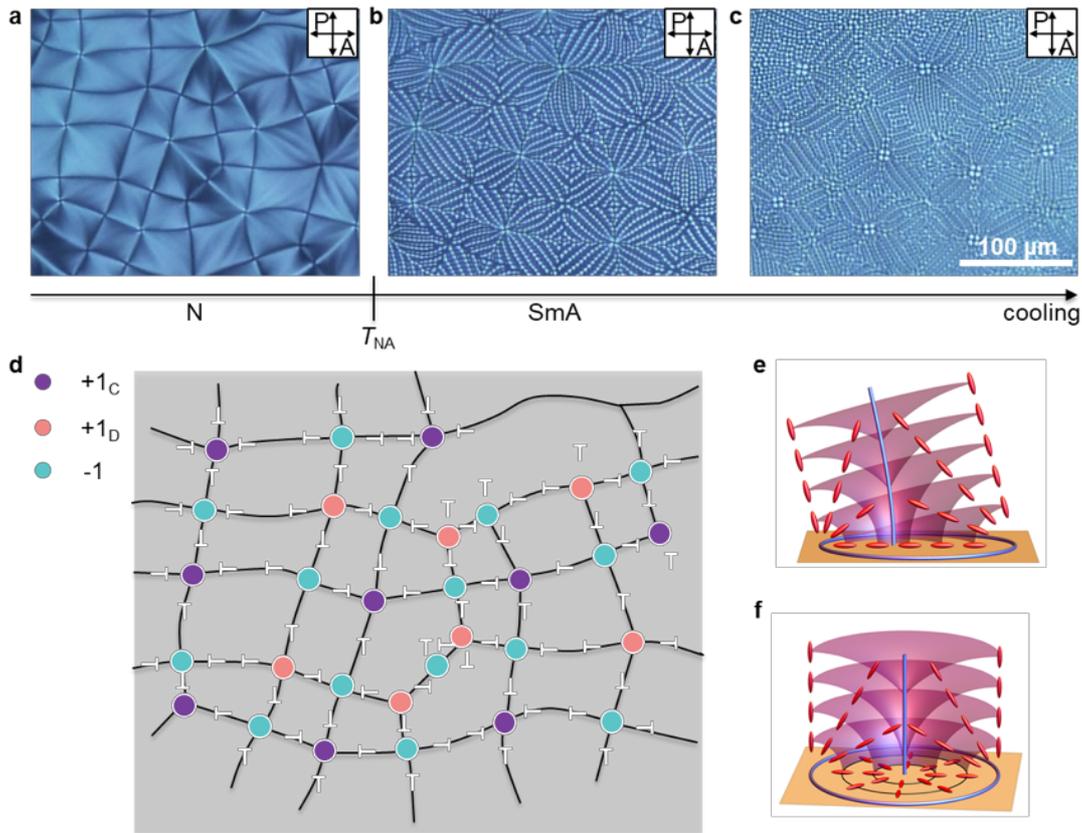

FIG. 2. Optical microscopic observation of ±1 boojums connecting continuously. (a) Spontaneously emerging boojums at N phase temperature. (b)-(c) FCDs with nonzero eccentricity at SmA phase temperature, (b) near the transition and (c) deeper in the SmA phase. (d) Schematic diagram of the director configuration corresponding to the schlieren texture of (a), using "nail-head" notation (top of "T" indicates side closer to viewer). The black lines indicate domain boundaries of large NLC domains created by $+1_C$ boojums. The purple dots, pink dots, and light blue dots correspond to $+1_C$ boojum, $+1_D$ boojum, and -1 boojum, respectively. (e,f) Illustrations of an FCD with nonzero eccentricity (e) and a TFCD with zero eccentricity (f), with Sm layers in purple, representative rod-like molecules in red, focal curves in blue, and the planar-anchoring substrate in orange.

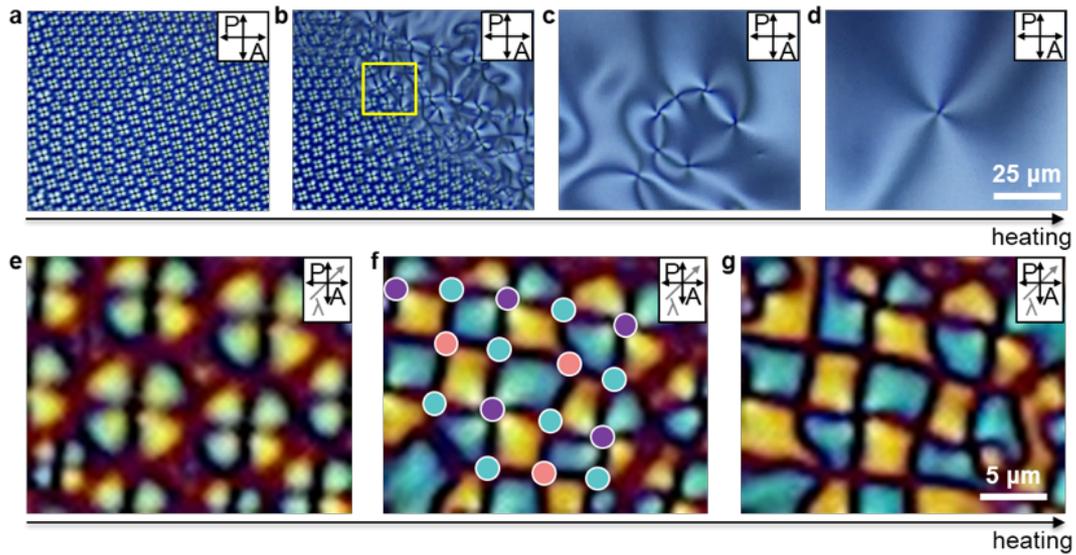

FIG. 3. Optical microscopic observation of TFCD-to-boojum transformation and annihilation of boojums during heating process from SmA to N phase. (a) Closely packed TFCDs at SmA phase. (b) Creation and annihilation of ±1 boojums at the vicinity of $T_{SmA-N}$. TFCDs transform to $+1_C$ boojums. $+1_D$ and $-1$ boojums form between TFCDs. (c)-(d) ±1 boojums annihilate to leave behind a single boojum in this region upon further heating. (e) Closely packed TFCDs at SmA phase. (f)-(g) TFCDs transform into $+1_C$ boojums (purple dots). Other boojums (pink dots: $+1_D$ boojums, light blue dots: $-1$ boojums) are generated between $+1_C$ boojums. The gray arrow indicates the slow axis of the first-order retardation plate. Cyan blue or yellow appears when the LC molecular slow axis is parallel or perpendicular to the slow axis of the first-order retardation plate, respectively. Magenta appears when the LC molecular slow axis is parallel, perpendicular, or vertical with respective to the polarizers.

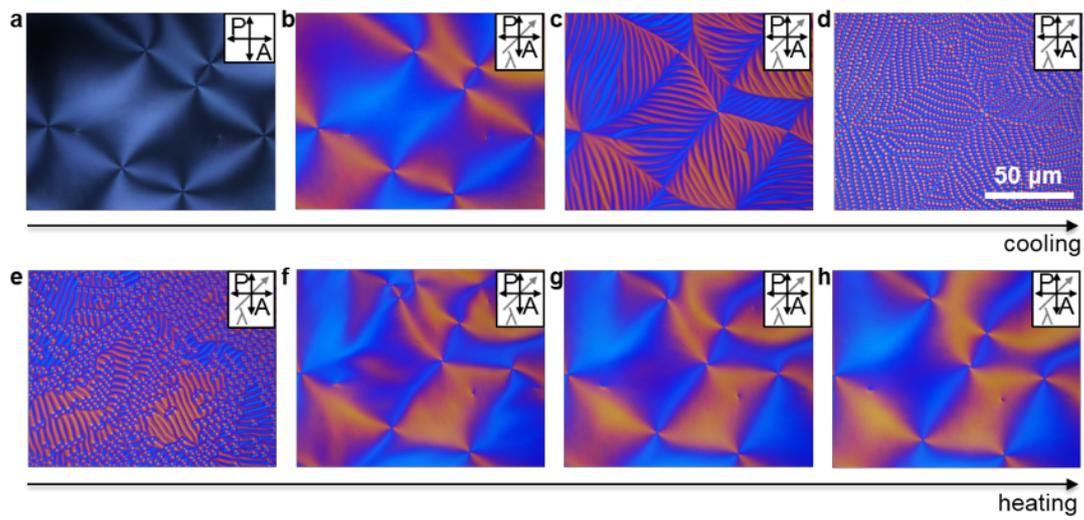

FIG. 4. Thermal reversibility of the defect configuration across the phase transition. (a)-(d) Cooling process. (d)-(h) Heating process. The positions of the defects are recovered after the formation and subsequent melting of FCDs in the SmA phase.

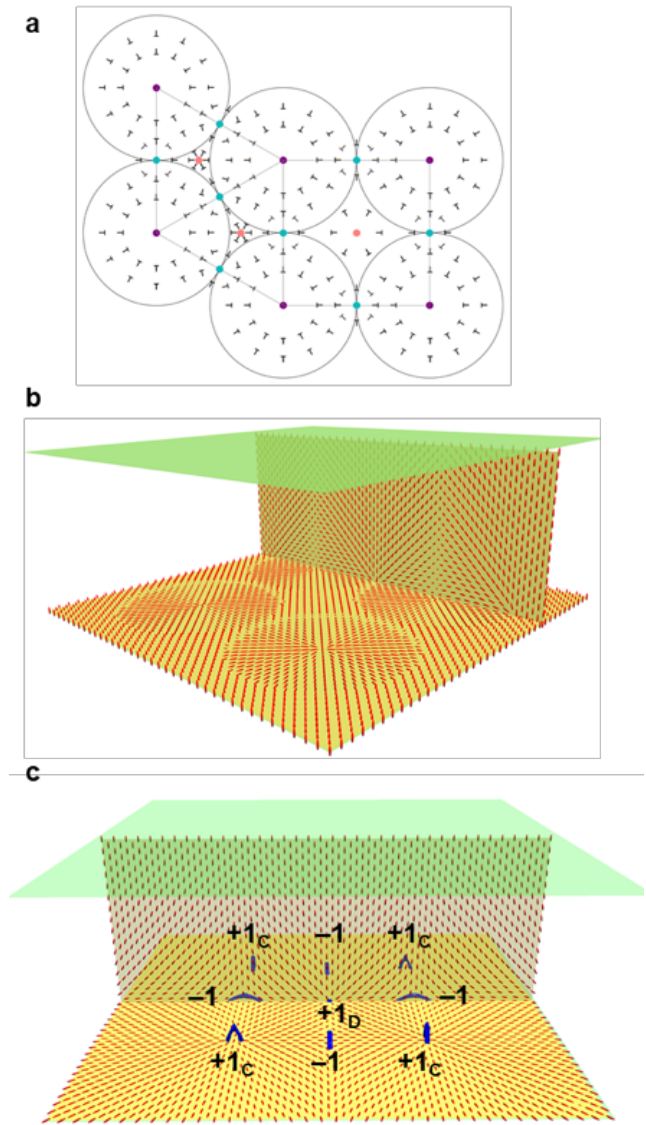

FIG. 5. Theoretical and numerical modeling of TFCD-to-boojum transformation. (a) Nail-head diagram of the hypothesized tilt directions *c* and boojums (purple: $+1_C$, pink: $+1_D$, light blue: $-1$) arising in the melting of an idealized TFCD packing. (b,c) Landau-de Gennes numerical modeling of the melting of a square of TFCDs. (b) Director field cross-section in the initial condition mimicking TFCDs. (c) Director field cross-section and boojums (blue) during energy relaxation. Four boojums are of type $+1_C$ and $-1$, and one is of type $+1_D$.

# Topological defects and geometric memory across the nematic-smectic A liquid crystal phase transition

*Ahram Suh[1], Min-Jun Gim[1], Daniel Beller[2]\* and Dong Ki Yoon[1,3]\**

[1]Graduate School of Nanoscience and Technology, KAIST, Daejeon, 34141,

Republic of Korea

E-mail: dbeller@ucmerced.edu (D.A.B.); nandk@kaist.ac.kr (D.K.Y.)

[2]Department of Physics, University of California, Merced, CA 95343, USA

[3]Department of Chemistry and KINC, KAIST, Daejeon, 34141, Republic of Korea

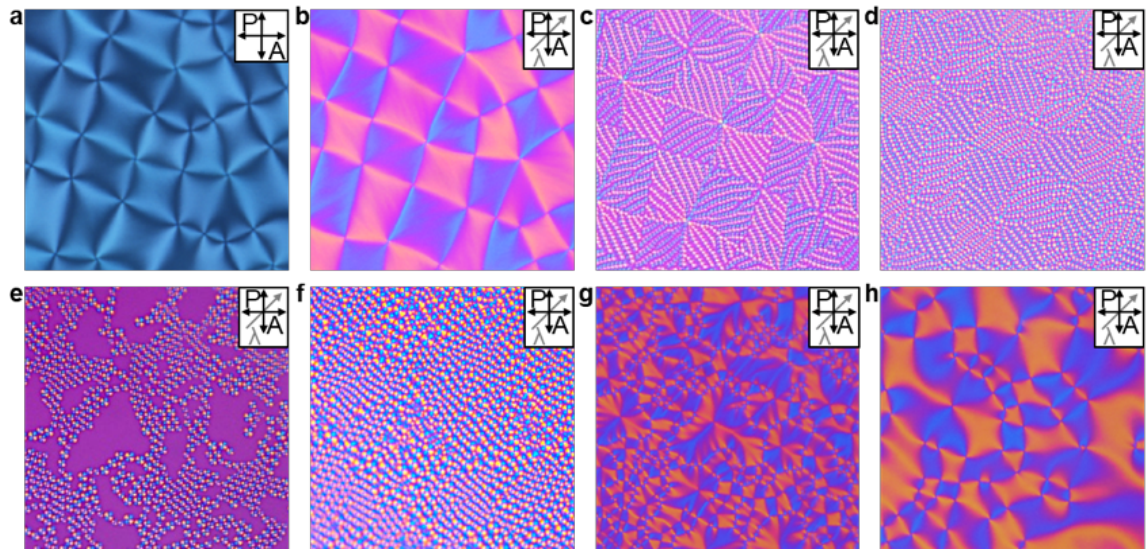

**FIG. S1. Optical microscopic observation of hysteresis of boojums during thermal process.** (a)-(e) Cooling process. (e)-(h) Heating process. Gray arrow indicates the slow axis of the inserted retardation plate, wherein cyan blue or yellow colors are shown when LC molecular long axis is parallel or perpendicular to the slow axis of the retardation plate, respectively. When the molecules orient parallel, perpendicular or vertical to polarizers, magenta is observed. Defects lose their position and configuration when they cooled to TFCDs at SmA phase.